\documentclass[print]{revtex4}
\usepackage{amsmath,amssymb}
\usepackage{graphicx}

\draft

\begin{document}

\title{Dynamics of Domain Wall in a Biaxial Ferromagnet With Spin-torque}

\author{Xin Liu\footnote{Electronic address:liuxin03@mail.nankai.edu.cn}, Xiong-Jun Liu
and Mo-Lin Ge} \affiliation{ Theoretical Physics Division, Nankai
Institute of Mathematics,Nankai University, Tianjin 300071,
P.R.China}

\begin{abstract}
The dynamics of the domain wall (DW) in a biaxial ferromagnet
interacting with a spin-polarized current are described by
sine-gordon (SG) equation coupled with Gilbert damping term in
this paper. Within our frame-work of this model, we obtain a
threshold of the current in the motion of a single DW with the
perturbation theory on kink soliton solution to the corresponding
ferromagnetic system, and the threshold is shown to be dependent
on the Gilbert damping term. Also, the motion properties of the DW
are discussed for the zero- and nonzero-damping cases, which shows
that our theory to describe the dynamics of the DW are
self-consistent.

 PACS numbers: 75.75.+a, 72.25.Ba, 75.30.Ds, 75.60.Ch
\end{abstract}
\maketitle

\baselineskip=15pt

\indent During the past decade or so, rapid advances have been
witnessed in both theoretical and experimental aspects toward
probing the novel mechanism of spin transfer torque, especially
focused on the motion of the domain wall (DW) induced by the
spin-polarized current.

Current-driven motion of a DW was studied in a series works by
Slonczewski \cite{1} and Berger \cite{2,3,4}. In particular, the
phenomenon that the electric current exerts a force on the DW via
the exchange coupling was argued in 1984 \cite{2}, and a
spin-polarized current exerts a torque on the wall magnetization
and also the motion due to the pulsed spin-polarized current are
studied in 1992 \cite{3}, etc.. In the following years these
interesting phenomena was observed by many
experiments\cite{5,6,7}. At the same time there are many
theoretical efforts \cite {th1,th2,th3,th4,th5,th6} to understand
the microscopic origins. The theoretical works, however, still
seem not satisfactorily to explain these phenomena and the
intrinsical reason is still unclearly. On the other hand, all
experiments \cite{5,6,7} indicate a threshold for the spin-current
which induces the motion of the DW, i.e. the motion doesn't happen
until the spin-current strength is lager than the value of this
threshold. So, a central theory of the induced domain-wall-motion
is to make clear the essential of the threshold, which we proceed
to study in this paper.

Very recently, the single-domain-wall as well as its dynamical
properties has been observed directly \cite{8}, and another
important phenomenon that current inducing a single DW switching
exists in ferromagnetic semiconductor has also been demonstrated
in experiment \cite{9}. All these experiments offer more
information for us to further understand it. At the same time,
many discussions have been presented in the refs
\cite{10,11,12,13}, especially the kind of spin-torque
\begin{eqnarray}\label{eqn:torque}
\tau=-b_{J}\frac{\partial \mathbf{M}}{\partial x}, \ \ \ \
b_{J}=\frac{\mu_{B}Pj_{e}}{eM_{s}}
\end{eqnarray}
was proposed firstly in \cite{10} and discussed in detail in the
refs. \cite{11,12}, where $\mathbf{M}$ is the ferromagnetic
magnetization, $M_{s}$ is the saturation magnetization, $P$ is the
spin polarization of the current, $\mu_{B}$ is the Bohr magneton
and $j_{e}$ is the electric current density. In this paper,
considering the above type of spin torque, we study the dynamical
properties of the DW in biaxial ferromagnet interacting with a
spin-polarized current, focusing on the threshold of the current
in the motion of the DW. The threshold is found to be dependent on
the Gilbert damping term from the stability condition of the DW's
static state, and the motion of the DW is shown to be a conclusion
with the minimal energy of the system.

Now, let us consider a spin-polarized current propagating in a
biaxial ferromagnetic material which has an easy x axis and a hard
z axis. Here we take an example of a Neel wall and treat the spin
configuration as uniform in y-z plane \cite{13}. We assume that
the length of magnetization $|\mathbf{M}|$ is constant and the
conducting electrons only interact with the local magnetization.
In this case, the modified Landau-Lifschitz equation reads
\begin{eqnarray}\label{lle}
\frac{\partial\mathbf{M}}{\partial t}=-\gamma
\mathbf{M}\times\mathbf{H}_{eff} + \frac{\alpha}{M_{s}} \mathbf{M}
\times \frac{\partial \mathbf{M}}{\partial t}+\tau
\end{eqnarray}
where $\gamma$ is the gyromagnetic ratio, \begin{eqnarray*}
\mathbf{H}_{eff}=\frac{2A}{M_{s}^{2}} \nabla^{2}
\mathbf{M}+\frac{H M_{x}}{M_{s}}e_{x}-\frac{(H_{\bot}+4\pi M_{s})
M_{z}}{M_{s}}e_{z}
\end{eqnarray*} is effective magnetic field which includes the exchange field,
anisotropy field and demagnetization field (note that there exists
no external field when the current propagates in the ferromagnet,
see, e.g. \cite{8,9}), H and $H_{\bot}$ are anisotropy fields
which correspond to easy axis x and hard axis z respectively, and
$\alpha$ is the Gilbert damping parameter. The spin-torque $\tau$
given by Eq. (\ref{eqn:torque}) only exists when the ferromagnet
has a DW structure such as a microfabricated magnetic wire. The
major object in present work is to study the threshold and the
dynamics of the DW with Eq. (\ref{lle}). To facilitate the further
discussion, we derive the above equation with spherical
coordinates $(\theta, \phi)$. By a straightforward calculation, we
reach the following equations of $\theta$ and $\phi$
\begin{eqnarray}\label{matrix}
{\left(
\begin{array}{lc}
\sin \theta  &   -\alpha\\
\alpha\sin \theta  & 1
\end{array}\right)}
{\dot{\phi} \choose \dot{\theta}}= {V_{\phi} \choose V_{\theta} }
\end{eqnarray}
where
\begin{eqnarray*}
V_{\phi}&=&-\frac{2A\gamma}{M_s}(\frac{\partial^{2}
\theta}{\partial x^{2}}-sin\theta cos\theta (\frac{\partial
\phi}{\partial x})^{2})-\gamma H sin\theta cos\theta sin\phi
cos\phi\nonumber\\
&&-\gamma(H_{\bot}+4\pi M_{s})sin\theta cos\theta- b_{J} sin\theta
\frac{\partial \phi}{\partial x}
\end{eqnarray*}
and
\begin{eqnarray*}
V_{\theta}=\frac{2A\gamma}{M_s}(sin\theta \frac{\partial^{2}
\phi}{\partial x^{2}}+2cos\theta \frac{\partial \theta}{\partial
x} \frac{\partial \phi}{\partial x})-\gamma H sin\theta sin\phi
cos\phi -b_{J} \frac{\partial \theta}{\partial x}
\end{eqnarray*}

For convenience we note $H_{\bot}+4\pi M_{s}$ as $H_{\bot}$ in the
following derivation. To obtain the threshold, we firstly
calculate the static solution to the Eq. (\ref{matrix}) and then
examine its stability condition with perturbation theory. This is
similar to the method used in the ref. \cite{10}. But here, the
magnetization $\mathbf{M}$ is a function of the space.

For the biaxial ferromagnet, one has $H_{\bot}\gg H, \sin\theta
\simeq 1$ and $\cos\theta\simeq 0$ \cite{integ}. Thus the static
solution is governed by $V_{\phi}=V_{\theta}=0$, i.e.
\begin{eqnarray}\label{V1}
b_{J}\frac{\partial\phi}{\partial x}+\gamma\cos\theta H_{\bot}=0,
\end{eqnarray}
\begin{eqnarray}\label{appro}
(\frac{2A\gamma}{M_s}-\frac{b_{J}^{2}}{\gamma H_{\bot}})
\frac{\partial^{2} \phi}{\partial x^{2}}-\gamma  H \sin\phi
\cos\phi=0.
\end{eqnarray}
With the boundary condition $\phi(-\infty)=0$ and
$\phi(\infty)=\pi$, one finds the form of the static kink soliton
\cite{perturbation} reads: $\phi=2\arctan[\exp
(\pm\frac{x}{\xi})]$, where these solutions are referred to
head-to-head DW (+ sign) and tail-to-tail DW (- sign) \cite{8} and
$\xi=\sqrt{\frac{2A}{MH}-\frac{b_{J}^2}{\gamma^{2}HH_{\bot}}}$
indicates that the spatial width of the soliton becomes narrower
when the current strength increases (Fig.1).
\begin{figure}[htbp]
\includegraphics[width=0.8\columnwidth]{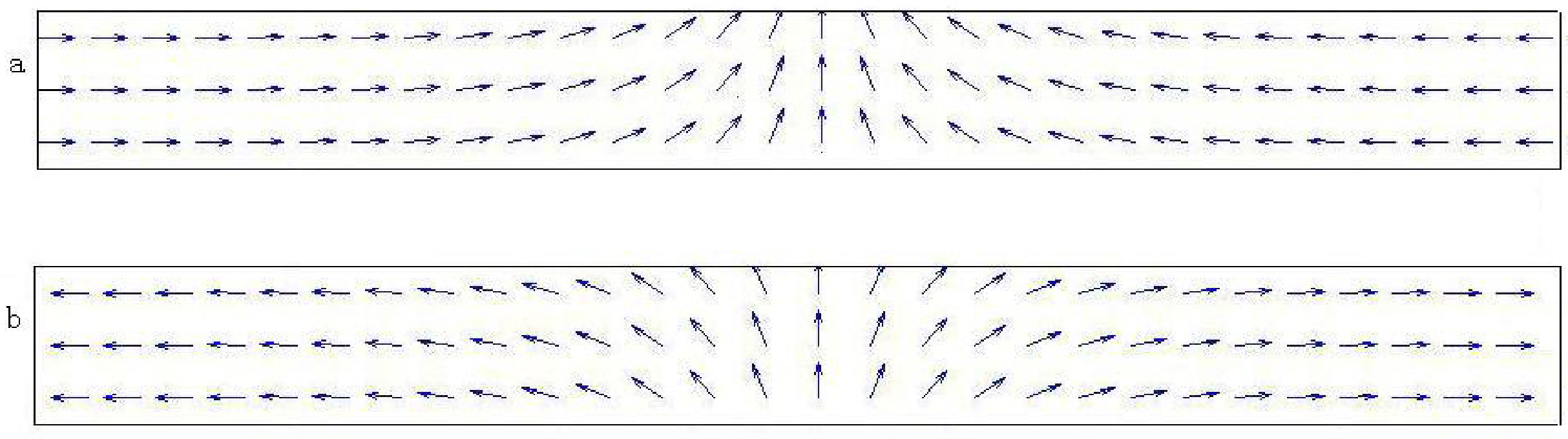}\\
\caption{(a) head-to-head DW, result of $\phi$=$2\arctan[\exp
(\frac{x}{\xi})]$. (b) tail-to-tail DW, result of
$\phi$=$2\arctan[\exp (-\frac{x}{\xi})]$.}\label{}
\end{figure}

In the following part, we consider the head-to-head DW and proceed
to calculate the threshold with perturbation theory. (This theory
can also be used in a tail-to-tail DW and gets the same result.)
For this we allow the static soliton solution develop a small
dynamic factor, i.e. $\phi\rightarrow\phi'$:
\begin{eqnarray}\label{perturb}
\phi'=2\arctan [\exp ((1+a(t))\frac{x}{\xi}+b(t))]
\end{eqnarray}
where $b(t)$ is a small perturbation and $a(t)$ is a higher order
smallness of $b(t)$ with $a(t)\sim(\frac{\partial b(t)}{\partial
t})^{2}$(This can be derived from the dynamic solution of SG
equation). Then, one has
$\delta\phi=\phi'-\phi=\frac{\partial\phi}{\partial x}b(t)$ and
the equations of $\delta\phi$, $\delta\theta$ can be written as
\begin{eqnarray}\label{eqn:6}
\frac{\partial}{\partial t}{\delta\phi \choose
\delta\theta}=\frac{1}{1+\alpha^{2}} {\left(
\begin{array}{lc}
-b_{J} \frac{\phi_{xx}}{\phi_{x}} &\gamma H_{\bot} \\
\alpha b_{J}\frac{\phi_{xx}}{\phi_{x}} &-\alpha \gamma H_{\bot}
\end{array} \right)}{\delta \phi \choose \delta \theta}=\frac{1}{1+\alpha^{2}}\hat{D} {\delta \phi \choose \delta
\theta}
\end{eqnarray}
One can verify that the eigenvalues $\lambda$ of the matrix $D$ is
given by
\begin{eqnarray*}
\lambda_{1}=0, \ \ \ \ \ \ \ \ \ \ \
\lambda_{2}=\frac{b_{J}}{\xi}\tanh(\frac{x}{\xi})-\alpha\gamma
H_{\bot}
\end{eqnarray*}

It is obvious that the stability condition of the solution $\phi$
depends on the properties of the eigenvalues \cite{stable}. Since
$\lambda_{1}$=0, the static solution can be steady only if
$\lambda_{2}\leq0$ at any position. For this one finds a critical
value for the current
\begin{eqnarray}\label{threshold}
J^{cr}_{e}=\frac{\alpha \gamma H_{\bot} e}{P \mu_{B}}
\sqrt{\frac{2A}{M_s H(1+\frac{\alpha^{2}H_{\bot}}{H})}}
\end{eqnarray}
above which the static DW can not be stable any more, i.e.
$J^{cr}_{e}$ is the threshold of the current. Since the value of
the function $\tanh x$ keeps approximate 1 from
$x\rightarrow\infty$ to the edge of the DW (we calculate this
value with matlab, e.g. when $x=2\xi$, one gets
$\tanh(2)=0.9640$), the DW becomes unstable firstly at its edge.
The pinning force \cite{13} is not considered in our derivation,
thus not the pinning force, but the Gilbert damping term $\alpha$
is the exclusive factor for existing a threshold in our argument.
Furthermore, if $\frac{\alpha^{2}H_{\bot}}{H}\ll 1$ (Because
$\alpha$ may be very small, it's not contrary with the assumption
that $H_{\bot}\gg H$), one finds the threshold is proportional to
the $H_{\bot}$ and the wall-parameter $\sqrt{\frac{2A}{M_s H}}$ as
well \cite{13}.

Now, we proceed to reveal properties of the motion for the DW
above the threshold. Also, we begin with Eq. (\ref{lle}). For the
case $H_{\bot}\gg H$, $\sin\theta \approx 1$ and $cos\theta
\approx 0$, the dynamic equation takes the following coupled form
\begin{eqnarray}\label{eqn:6}
\sin\theta \frac{\partial \phi}{\partial t} -\alpha \frac{\partial
\theta}{\partial t}=-b_{J} \frac{\partial
\phi}{\partial x}-\gamma cos\theta H_{\bot}
\end{eqnarray}
\begin{eqnarray}\label{eqn:coupled}
\alpha sin\theta \frac{\partial \phi}{\partial t}+\frac{\partial
\theta}{\partial t}= \frac{2A\gamma}{M_s} \frac{\partial^{2}
\phi}{\partial x^{2}}-\gamma H sin\phi cos\phi-b_{J}\frac{\partial
\theta}{\partial x}
\end{eqnarray}

As the method used in \cite{12}, we differentiate equation
(\ref{eqn:6}) and substitute it into Eq. (\ref{eqn:coupled})
yields
\begin{eqnarray}\label{eqn:decouple}
\frac{\partial^{2} \phi}{\partial t^{2}} -(\frac{2A \gamma^{2}
H_{\bot}}{M_s}-b_{J}^{2}) \frac{\partial^{2} \phi}{\partial x^{2}}
+ 2b_{J} \frac{\partial^{2} \phi}{\partial x
\partial t}+\gamma^{2}H H_{\bot} sin\phi cos\phi+\alpha \gamma
H_{\bot}\frac{\partial \phi}{\partial t}=0
\end{eqnarray}
From above formula one finds an additional term $2b_{J}\phi_{xt}$
(the third term of the left side) of Eq. (\ref{eqn:decouple}) is
added due to the spin current. To facilitate the further
discussions, we make transformations: $t^{'}=t cos\beta+\kappa x
sin\beta, \ \ x^{'}=-tsin\beta+\kappa x cos\beta, \
\tan2\beta=b_{J}\kappa$, where $\kappa=1/\sqrt{\frac{2A \gamma^{2}
H_{\bot}}{M_s}-b_{J}^{2}}=\sqrt{\frac{M_s}{2A\gamma^{2}H_{\bot}}}\zeta$
with $\zeta>1$. Using these new coordinates, the Eq.
(\ref{eqn:decouple}) recasts into
\begin{eqnarray}\label{eqn:8}
\frac{1}{\cos 2\beta}(\frac{\partial^{2} \psi}{\partial
t^{'2}}-\frac{\partial^{2} \psi}{\partial x^{'2}}) +\gamma^{2}H
H_{\bot}sin\psi+\alpha \gamma H_{\bot}(cos\beta\frac{\partial
\psi}{\partial t^{'}}-sin\beta\frac{\partial \psi}{\partial
x^{'}})=0
\end{eqnarray}
where $\psi=2\phi$. To give a more intuitive discussion, we
neglect the damping term in the following derivation. Then, under
the boundary condition: $\psi(-\infty)=0 \ \ \psi(\infty)=2\pi$,
we obtain kink soliton solution of the Eq. (\ref{eqn:decouple}):
\begin{eqnarray}\label{eqn:9}
\phi&=&2arctan[\exp(\sqrt{\frac{\gamma^{2}H H_{\bot}\cos
2\beta}{(1-v^{2})}}(x^{'}-vt^{'}))]\nonumber\\
&&=2arctan[\exp(\sqrt{\frac{\gamma^{2}H H_{\bot}\cos
2\beta}{(1-v^{2})}}(\kappa x
(\cos\beta-v\sin\beta)-t(\sin\beta+v\cos\beta)))]
\end{eqnarray}
where $v$ is the velocity of the DW in $x^{'}-t^{'}$ coordinate
and a free parameter subject only to the restriction $|v|\leq1$,
and it is related to the velocity $\mathbf{V}$ by
$\mathbf{V}$=$\frac{v\cos \beta+\sin \beta}{\kappa(\cos\beta-v\sin
\beta)}$. With this solution we can calculate the energy of the DW
for the motion case. Deriving from the Lagrangian $L=\int d^{3}x
(M_s \dot{\phi}(\cos \theta -1)-E)$, one can obtain the
hamiltonian in the following form:
\begin{eqnarray}\label{H}
H=\int d^{3}x E=\int d^{3}x
(2A((\nabla\theta)^{2}+\sin\theta(\nabla\phi)^{2})-H
M_s(\sin\theta \cos\phi)^{2}+H_{\bot}M_s\cos^{2}\theta).
\end{eqnarray}

This result is valid no matter whether the DW is steady or not,
and it indicates that there is no kinetic energy in the
hamiltonian even for the propagating case of the domain wall.
Substituting the Eq. (\ref{eqn:9}) into (\ref{H}), we obtain the
DW energy $E$ as
\begin{eqnarray}\label{energy}
E&=&2\sqrt{2AM_s H\cos2\beta}\zeta
\{\frac{\cos\beta-v\sin\beta}{\sqrt{1-v^{2}}}+\frac{
\sqrt{1-v^{2}}}{\cos2\beta\zeta^{2}(\cos\beta-v\sin\beta)}+\nonumber\\
&&\frac{M_s}{\gamma^{2}H_{\bot}2A} \frac{(b_{J}
(\cos\beta-v\sin\beta)-(\sin\beta+v\cos\beta)/\kappa)^{2}}{\sqrt{1-v^{2}}(\cos\beta-v\sin\beta)}
\}+C
\end{eqnarray}
where $C$ is a constant independent on any parameters. Eq.
(\ref{energy}) tells us that the total energy of the DW is the
function of the parameters $v$ and $\beta$. The first term in the
bracket of the right side of above equation corresponds to
exchange energy, the second term corresponds to the anisotropy
energy of the easy axis and the third one corresponds to the
anisotropy energy of the hard axis. By a straightforward
calculation we find that the total energy E reaches its minimum
when $v=\tan\beta$. \begin{figure}
  \includegraphics[width=0.5\columnwidth]{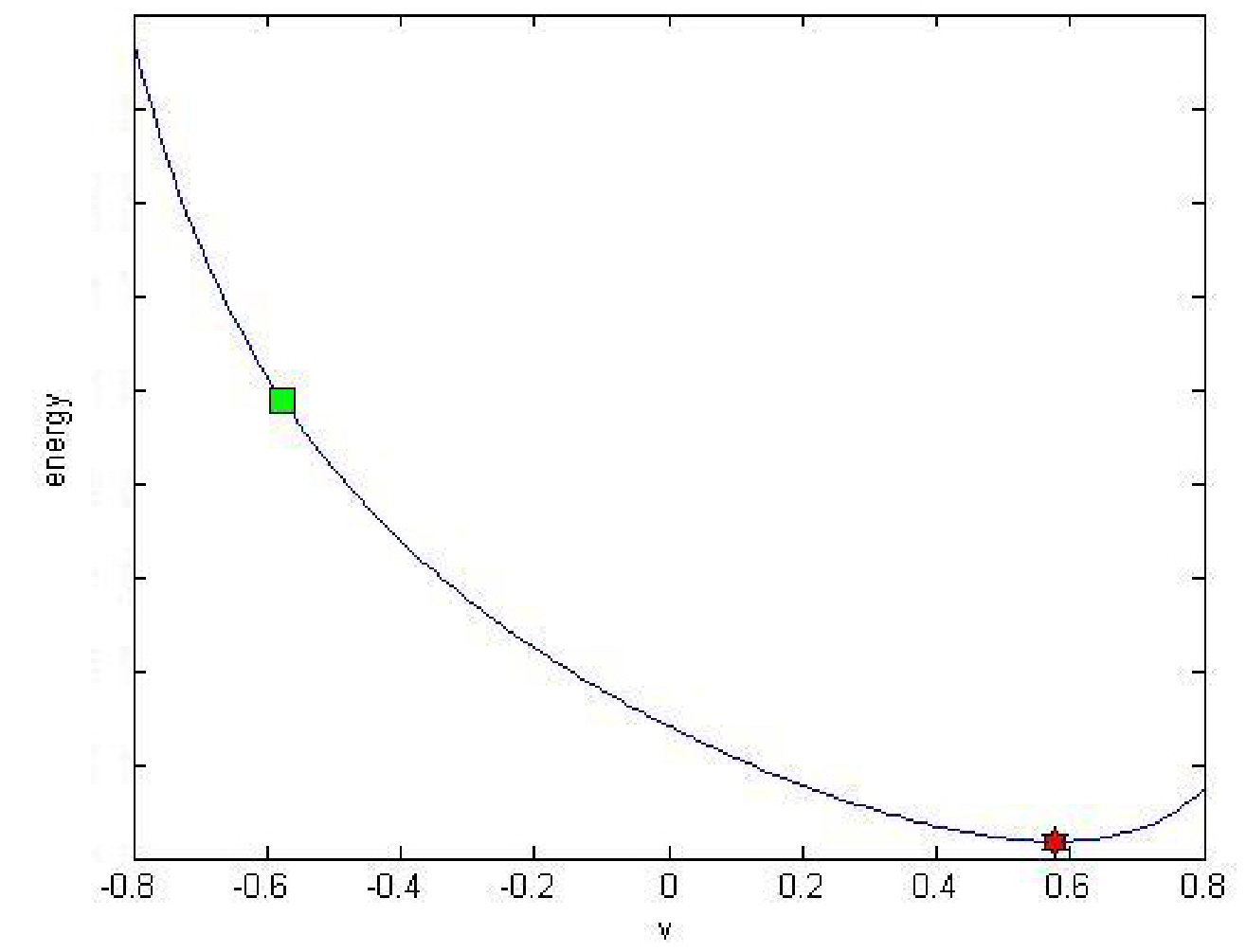}\\
  \caption{DW energy as a function of parameter $v$($|v|\leq 1$).
  The velocity of the DW $\mathbf{V}$=$\frac{v\cos \beta+\sin \beta}{\kappa(\cos\beta-v\sin
\beta)}$. Green square represents the energy of DW when
$\mathbf{V}=0$ and red hexagon
   represents the minimal energy of DW when $\mathbf{V}=b_{J}$(we take an example of $\beta=\frac{\pi}{6}$).}\label{}
\end{figure}
Therefore in this case, the DW structure can be described
as
\begin{eqnarray}\label{min}
\phi_{min}=2\arctan[\exp(\frac{\cos 2\beta}{\xi}(x-b_{J}t))]
\end{eqnarray}
The above formula clearly shows that the distribution of
$\phi_{min}$ is similar to the static soliton solution except for
a velocity $b_{J}$ and $M_{z}=0$. So, if the damping term
$\alpha=0$, the DW moves once the spin current is applied, and the
threshold arrives at zero. These results are self-consistent with
the former derivation that the threshold is proportional to the
damping $\alpha$ (see eq. (\ref{threshold})).

Before sum up, we should point out that practically the velocity
observed in experiment will be smaller than $b_{J}$ for the
nonzero damping term. For a brief discussion on this effect one
may assume the initial velocity of the DW is $b_{J}$ \cite{12} in
the above-threshold case. With perturbation theory on the SG
soliton, after Lorentz transformation to the soliton's rest frame
$z-\tau$ and noting that $\psi=2\phi$, we obtain a series of
equations similar to that in ref. \cite{perturbation}:
\begin{eqnarray}\label{dsoliton}
\phi(z,\tau)=\phi_{min}(z,\tau)+\xi(z,\tau),
\end{eqnarray}
here $\phi_{min}$ (see eq. (\ref{min})) is described in $z-\tau$
frame and $|\xi(z,\tau)|\ll 1$ \cite{perturbation} and
\begin{eqnarray}\label{sum}
\xi(z,\tau)=\frac{1}{8}u_{b}(\tau)f_{b}(z)+\int^{\infty}_{-\infty}
dk u_{k}(\tau)f_{k}(z),
\end{eqnarray}
\begin{eqnarray}\label{rest} \frac{\partial^{2} \psi}{\partial
\tau^{2}}-\frac{\partial^{2} \psi}{\partial z^{2}}
+\sin\psi+\nu(\frac{\partial \psi}{\partial
\tau}-\tan2\beta\frac{\partial \psi}{\partial z})=0,
\end{eqnarray}
where $f_{b}(z)=2sech(z)$,
$f_{k}(z)=(2\pi)^{\frac{1}{2}}\omega_{k}^{-1}[k+i\tanh(z)]\exp(ikz)
$, $\omega_{0}^{2}$=$\gamma^{2}H H_{\bot}\cos2\beta$,
$\omega_{k}=1+k^{2}$ and $\nu$=$\frac{\alpha \gamma
H_{\bot}\cos^{\frac{3}{2}}2\beta}{\omega_{0}}$. Note that in Eq.
(\ref{rest}) $\tan2\beta$ is not a small value, we approximately
obtain the amplitude of the translational mode $u_{b}$ satisfies
\begin{eqnarray}\label{co}
\frac{du_{b}(\tau)}{d\tau}=(8\tan2\beta(1-\exp(-\nu\tau))+\exp(-\nu\tau)\int
f(\tau)\exp(\nu\tau) d\tau)
\end{eqnarray}
with
$f(\tau)=i(\pi/2)^{\frac{1}{2}}(\tan2\beta)\nu\int^{\infty}_{-\infty}
dk [k u_{k}(\tau)]/(\omega_{k}\sinh \frac{1}{2}\pi k)$. This
formula means that in $z-\tau$ frame, the DW will get a velocity
equal to $(-du_b/d\tau)$ \cite{perturbation}. From above result
one easily find the velocity of the DW $\mathbf{V}$ should be
smaller than $b_{J}$. While we can not let $\tau\sim\infty$ to get
the terminal velocity because for large $\tau$, this perturbation
theory breaks down since $u_{b}$ grows with $\tau$. In fact,
theoretically the motion will not stop as long as the spin current
strength is larger than the threshold value, because the damping
term will disappear once the velocity becomes zero and then
according to present result this static solution is obviously a
high energy state, which is not steady.

The development herein is outlined as follows. Initially, no spin
current is applied to the biaxial ferromagnet, the variable
$\theta=\pi/2$ and the DW can be described as static soliton
solution of SG equation. Secondly, a current is applied while the
DW still keeps static if the current strength is smaller than the
threshold value, and its width becomes short in x-y plane. On the
other hand, the magnetization $\mathbf{M}$ develops a component
$M_{z}$ as is shown in Eq. (\ref{V1}),
\begin{figure}[htbp]
  \setlength{\belowcaptionskip}{0pt}
  \includegraphics[width=1.0\columnwidth]{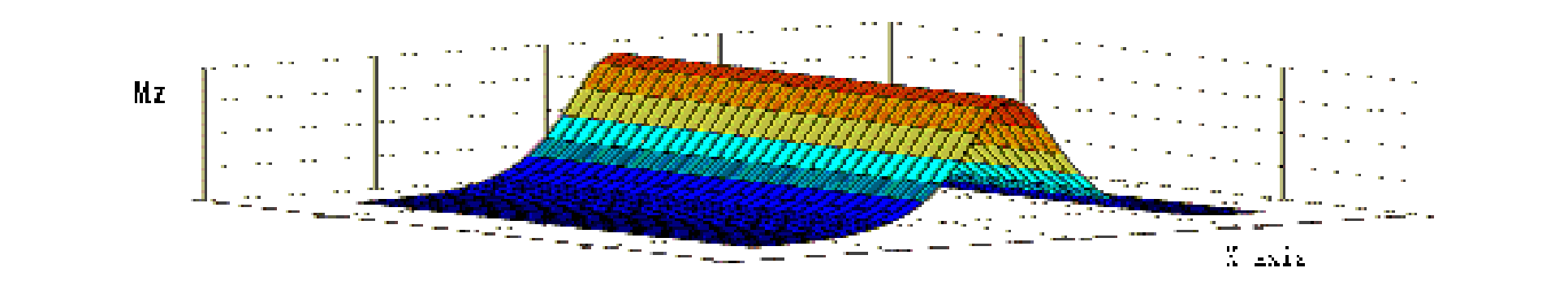}\\
  \caption{With spin-torque, the out-of-plane component $M_{z}$ as a function of $x$-coordinate,
   is uniform in y-z plane. }\label{}
\end{figure}the value of
$M_{z}$ and the anisotropy energy of hard axis increase with time
due to the increasing current strength till which arrives at the
threshold (Fig. 3). Thirdly, for the above-threshold case, the
static DW can't support the increasing of $M_{z}$, i.e. this case
will result in the propagation of the DW along the motion
direction of the conducting electrons in the current. Then,
neglecting the Gilbert damping term, we discuss the dynamical
properties of Landau-Lifshitz equation by discovering a series of
solutions which respectively correspond to different energies of
the DW and find the velocity of the DW induced by the spin-current
is $b_J$, which corresponds to the minimal energy of the DW.
Finally, for the nonzero-damping case, the velocity of the DW is
found to be smaller than $b_{J}$. With these derivations we show
our result is self-consistent that the threshold depends on the
Gilbert damping term of the coupled Landau-Lifschitz equation.

In conclusion the dynamical properties of a single DW is studied
in biaxial ferromagnet induced by a torque exerted via the
spin-polarized current. With the perturbation theory on the static
kink soliton solution to the ferromagnetic system, we obtain a
threshold of the current in the motion of domain wall and the
threshold is shown to be dependent on the Gilbert damping term of
the coupled Landau-Lifschitz equation. Also, the motion properties
of the DW are discussed in the zero- and nonzero-damping cases,
which shows that our theory to describe the dynamics of the DW are
self-consistent.

The authors are grateful to B. S. Han and D. J. Zheng for their
wonderful lectures on spintronics in Nankai University. It's these
lectures that lead us to this attractive field. This work is
supported by NSF of China under grants No.10275036 and
No.10304020.



\noindent\\  \\

\end{document}